\documentclass[
  aps,
  prl,
  reprint,
  twocolumn,
  superscriptaddress,
  floatfix,
  nofootinbib,
  amsmath,
  amssymb
]{revtex4-2}

\usepackage{graphicx}
\usepackage{siunitx}
\usepackage[dvipsnames]{xcolor}
\usepackage[colorlinks=true,linkcolor=blue!50!black,citecolor=blue!50!black,urlcolor=blue!50!black]{hyperref}

\begin{document}

\title{FPGA-based disturbance-observer servo for broadband noise suppression in laser frequency stabilization}

\author{Meung Ho Seo}
\affiliation{Time and Frequency Group, Korea Research Institute of Standards and Science, Daejeon 34113, Republic of Korea}

\author{Jae Hoon Lee}
\affiliation{Atomic Quantum Sensing Group, Korea Research Institute of Standards and Science, Daejeon 34113, Republic of Korea}

\author{Young-Ho Park}
\affiliation{Time and Frequency Group, Korea Research Institute of Standards and Science, Daejeon 34113, Republic of Korea}

\author{Hyun-Gue Hong}
\affiliation{Time and Frequency Group, Korea Research Institute of Standards and Science, Daejeon 34113, Republic of Korea}

\author{Seji Kang}
\affiliation{Time and Frequency Group, Korea Research Institute of Standards and Science, Daejeon 34113, Republic of Korea}

\author{Myoung-Sun Heo}
\affiliation{Time and Frequency Group, Korea Research Institute of Standards and Science, Daejeon 34113, Republic of Korea}

\author{Sang-Bum Lee}
\affiliation{Atomic Quantum Sensing Group, Korea Research Institute of Standards and Science, Daejeon 34113, Republic of Korea}

\author{Taeg Yong Kwon}
\affiliation{Atomic Quantum Sensing Group, Korea Research Institute of Standards and Science, Daejeon 34113, Republic of Korea}

\author{Sangwon Seo}
\affiliation{Atomic Quantum Sensing Group, Korea Research Institute of Standards and Science, Daejeon 34113, Republic of Korea}

\author{Sang Eon Park}
\email{parkse@kriss.re.kr}
\affiliation{Time and Frequency Group, Korea Research Institute of Standards and Science, Daejeon 34113, Republic of Korea}

\date{\today}

\begin{abstract}
We have demonstrated broadband frequency-noise suppression in a laser stabilization system by augmenting a conventional proportional-integral-derivative (PID) controller with a digital disturbance observer (DOB) implemented on a field-programmable gate array (FPGA).
The DOB employs a first-order exponential moving average filter as its Q-filter, replacing multi-parameter frequency-domain plant identification with a single one-dimensional gain sweep.
Using modulation transfer spectroscopy on the $^{87}$Rb D$_2$ line at \SI{780.24}{\nano\meter}, we have measured the frequency-noise power spectral density and the Allan deviation of the beat note between two independently stabilized lasers.
The integrated rms frequency noise below \SI{40}{\kilo\hertz} decreased by \SI{16.9}{\decibel} compared with PID alone, corresponding to a reduction from approximately \SI{140}{\kilo\hertz} to \SI{20}{\kilo\hertz}.
The short-term fractional frequency instability improved from $\sigma_y(\SI{1}{\milli\second}) = 7.9 \times 10^{-12}$ to $4.6 \times 10^{-12}$, while the long-term stability at $\tau > \SI{1}{\second}$ remained within statistical uncertainty.
This DOB-augmented architecture offers a simple and effective route to enhanced noise rejection in FPGA-based servo systems for atomic physics experiments.
\end{abstract}

\maketitle

\section{Introduction}

Precise and reliable laser frequency stabilization is an essential prerequisite for a wide range of experiments in atomic, molecular, and optical (AMO) physics.
Optical atomic clocks~\cite{Ludlow2015}, atomic fountain clocks~\cite{Wynands2005, Bize2005}, atom interferometers~\cite{Kasevich1991}, quantum sensors~\cite{Bongs2019}, and trapped-ion quantum processors~\cite{Bruzewicz2019} all rely on laser sources whose frequency noise must be suppressed to levels dictated by the specific application requirements.
As these technologies mature toward field-deployable instruments---including transportable optical clocks~\cite{Grotti2018}, quantum gravimeters~\cite{Menoret2018}, space-borne atom interferometers~\cite{Becker2018}, compact satellite references~\cite{Strangfeld2022}, and portable atom-interferometry laser systems~\cite{Schmidt2011}---the demand for compact, robust, and high-performance laser stabilization systems has grown significantly.

The standard approach to laser frequency stabilization is the proportional-integral-derivative (PID) controller~\cite{Bechhoefer2005}, typically implemented either in analog electronics or, more recently, on field-programmable gate arrays (FPGAs)~\cite{Preuschoff2020, Wiegand2022, Neuhaus2024, Avalos2023, Leibrandt2015, Appel2009}.
While PID controllers are well understood and straightforward to tune, it is difficult to achieve sufficient loop gain at intermediate frequencies---above the unity-gain bandwidth of the integrator but below the closed-loop servo bandwidth---without compromising phase margin.
External disturbances in this frequency range, arising from acoustic vibrations, mechanical resonances, or electronic pickup, can degrade the frequency noise power spectral density (PSD)~\cite{DiDomenico2010, Numata2004} and consequently the short-term frequency instability.

The disturbance observer (DOB) is a robust control technique introduced by Ohnishi and colleagues for motion control of mechatronic systems~\cite{Ohnishi1996}.
Subsequently formalized within a two-degree-of-freedom control framework by Umeno and Hori~\cite{Umeno1991}, the DOB has become a standard tool in motion control~\cite{Kempf1999, Schrijver2002}, robotics, and precision positioning~\cite{Sariyildiz2014}, as reviewed comprehensively by Chen \textit{et al.}~\cite{Chen2016DOBReview} and Sariyildiz \textit{et al.}~\cite{Sariyildiz2020}.
The essential idea is to estimate the lumped disturbance acting on a plant by comparing the measured output with the output predicted from a nominal plant model, filtering the difference through a low-pass Q-filter, and feeding back the filtered estimate to cancel the disturbance.
Crucially, Shim and Jo~\cite{Shim2009} have shown that robust stability can be guaranteed even when the nominal plant model is only approximate, provided that the Q-filter bandwidth is chosen conservatively.
This property makes DOB particularly attractive for laser frequency stabilization, where the plant transfer function (from control voltage to optical frequency) is difficult to identify precisely due to the interplay of current modulation, thermal effects, and mechanical compliance of the laser cavity.

Although related model-based and feed-forward approaches have been applied to adjacent precision-control problems~\cite{Chao2024, Li2023}, an explicit Q-filter DOB has not been widely deployed in optical frequency stabilization. The present work introduces such a DOB with the inverse plant model reduced to a single scalar gain $K_n$ and an FPGA bit-shift implementation that uses no DSP slices.
In this work, we have implemented a digital DOB alongside a PID controller on a Red Pitaya STEMlab 125-14 LN FPGA platform, operating at a clock frequency of \SI{125}{\mega\hertz}.
The Q-filter is realized as a simple first-order exponential moving average (EMA) filter, requiring only a single tunable parameter (the cutoff frequency) and no prior identification of the plant transfer function.
Using modulation transfer spectroscopy (MTS) on the $^{87}$Rb D$_2$ transition at \SI{780.24}{\nano\meter}~\cite{Shirley1982, McCarron2008, Noh2011}, we have characterized the frequency-noise power spectral density (PSD) and the overlapping Allan deviation~\cite{Allan1966, Allan1987} of the beat note between two independently stabilized laser systems.
We have found that the DOB reduces the integrated rms frequency noise below \SI{40}{\kilo\hertz} by approximately \SI{17}{\decibel} compared with PID alone, confirming the low-pass nature of the disturbance rejection.

\section{Theory}

The feedback system considered here consists of two controllers acting on the error signal $e(t)$: a conventional PID controller and a DOB.
We first describe the operating principle of each in continuous-time, then discuss the frequency-domain properties of the combined system.

A PID controller produces a control signal $u_\text{PID}$ from the error signal $e(t) = r(t) - y(t)$, where $r(t)$ is the setpoint and $y(t)$ is the measured output.
In the Laplace domain, the controller transfer function is
\begin{equation}
 C(s) = K_p + \frac{K_i}{s} + K_d\, s\,,
 \label{eq:pid}
\end{equation}
where $K_p$, $K_i$, and $K_d$ are the proportional, integral, and derivative gains, respectively.
The proportional term responds instantaneously to the current error, the integral term accumulates past errors to eliminate steady-state offset, and the derivative term acts on the rate of change to provide phase lead.
At low frequencies the integrator dominates the loop gain and is primarily responsible for eliminating steady-state offset, while the proportional and derivative terms extend the servo bandwidth to higher frequencies.
However, the achievable PID loop gain at intermediate frequencies is limited by the requirement to maintain adequate phase margin, so external disturbances at these frequencies are only partially suppressed.

\begin{figure}[!t]
\centering
\includegraphics[width=\columnwidth]{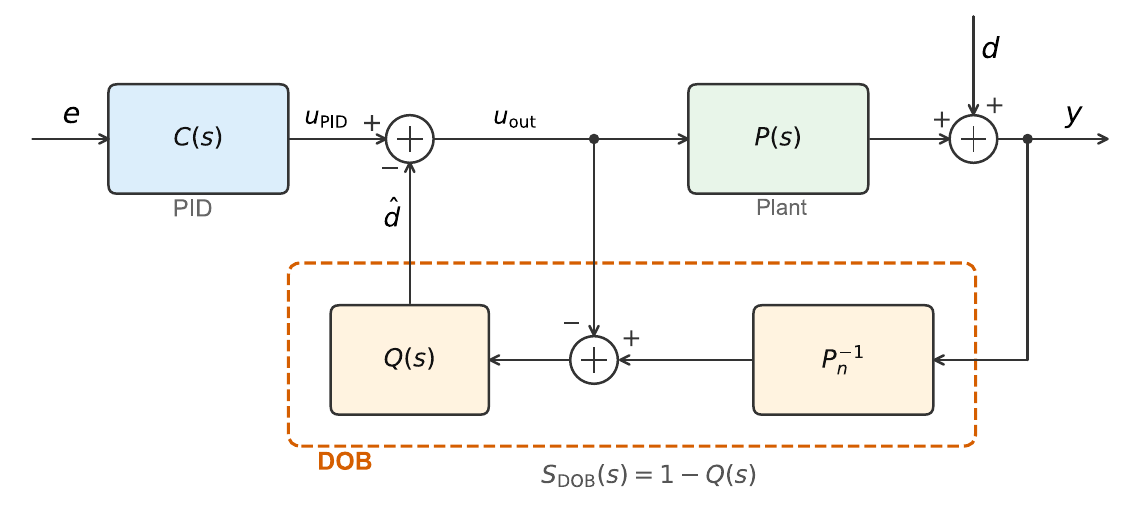}
\caption{Block diagram of the PID+DOB control architecture. The PID controller $C(s)$ generates the control signal $u_\text{PID}$ from the error $e$. The plant $P(s)$ converts the input $u_\text{out}$ to the output $y$, which is corrupted by the disturbance $d$. The DOB inner loop (dashed box) estimates $d$ by passing $P_n^{-1} y - u_\text{out}$ through the Q-filter $Q(s)$, and subtracts the estimate $\hat{d}$ from the control signal. The DOB sensitivity function $S_\text{DOB}(s) = 1 - Q(s)$ determines the low-pass disturbance rejection.}
\label{fig:dob_block}
\end{figure}

The disturbance observer addresses this limitation by adding a parallel inner loop that estimates and compensates for the lumped disturbance $d(t)$ acting on the plant.
Consider the control system shown in Fig.~\ref{fig:dob_block}: the PID controller $C(s)$ produces the nominal control signal $u_\text{PID}$, which is combined with the DOB correction to form the plant input $u_\text{out}$.
The plant $P(s)$ converts $u_\text{out}$ into the output $y(t)$, which is corrupted by the lumped disturbance $d(t)$:
\begin{equation}
 Y(s) = P(s)\, U_\text{out}(s) + D(s)\,,
 \label{eq:plant}
\end{equation}
where $Y(s)$, $U_\text{out}(s)$, and $D(s)$ are the Laplace transforms of the output, control input, and disturbance, respectively.

The DOB estimates $d(t)$ using a nominal model $P_n(s)$ of the plant.
If $P_n(s)$ were an exact replica of $P(s)$, one could recover the disturbance simply as $\hat{D} = P_n^{-1} Y - U_\text{out}$.
In practice, however, $P_n^{-1}(s)$ may be improper (more zeros than poles), and high-frequency noise would be amplified.
The Q-filter $Q(s)$ resolves this by low-pass filtering the disturbance estimate:
\begin{equation}
 \hat{D}(s) = Q(s)\bigl[\,P_n^{-1}(s)\, Y(s) - U_\text{out}(s)\,\bigr]\,.
 \label{eq:d_est}
\end{equation}
The compensated control signal is then
\begin{equation}
 U_\text{out}(s) = U_\text{PID}(s) - \hat{D}(s)\,,
 \label{eq:uout}
\end{equation}
which closes the inner disturbance-rejection loop around the PID controller.

The key design parameter is the Q-filter, which determines the frequency range over which disturbances are rejected.
A first-order low-pass filter with cutoff frequency $\omega_c$ is a convenient choice---unity gain at DC, monotonic roll-off, and a single tunable parameter:
\begin{equation}
 Q(s) = \frac{\omega_c}{s + \omega_c}\,,
 \label{eq:qfilter}
\end{equation}
with $|Q(j\omega)| = 1$ at DC and rolling off to zero above $\omega_c$.
Assuming a perfect nominal model ($P_n = P$), the closed-loop transfer function from disturbance $d$ to output $y$ in the presence of the DOB can be expressed via the sensitivity function
\begin{equation}
 S_\text{DOB}(s) = 1 - Q(s)\,.
 \label{eq:sdob}
\end{equation}
This function fully characterizes the DOB's disturbance rejection.
Defining the cutoff frequency as $f_c = \omega_c / 2\pi$, at frequencies $f \ll f_c$, $|Q| \approx 1$ and $|S_\text{DOB}| \ll 1$, so disturbances are strongly suppressed. At $f \gg f_c$, $|Q| \approx 0$ and $|S_\text{DOB}| \approx 1$, so the DOB becomes transparent and does not interfere with the PID controller.
This frequency-selective behavior is the central advantage of the DOB---it adds noise rejection at frequencies below the Q-filter cutoff $f_c$, where PID gain alone is limited, while leaving the high-frequency response unchanged. The DOB thus functions as a low-pass augmentation of the existing PID loop.

A further important property concerns the choice of the nominal plant model $P_n(s)$.
In many practical situations the exact plant transfer function is difficult to identify.
However, the Q-filter inherently limits the DOB's action to low frequencies, so the model $P_n$ need only be accurate where $Q(s)$ is significant.
If the plant can be approximated as a simple gain at frequencies below $f_c$, the inverse model reduces to a scalar multiplication $P_n^{-1}(s) \approx K_n$, and Eq.~(\ref{eq:d_est}) simplifies to
\begin{equation}
 \hat{D}(s) = Q(s)\bigl[\,K_n\, Y(s) - U_\text{out}(s)\,\bigr]\,.
 \label{eq:d_est_simple}
\end{equation}
The essential DOB structure---Q-filtered estimation of the lumped disturbance---and the sensitivity function $S_\text{DOB}(s) = 1 - Q(s)$ remain approximately valid under this simplification, at frequencies where $P(s) \approx P_0$.

The PID+DOB architecture maps efficiently onto digital hardware.
In a sampled-data system with clock frequency $f_\text{clk}$, the PID controller is implemented as a parallel sum of proportional, integral (accumulator), and derivative (first difference) paths---a standard digital realization requiring only multiplications and additions per clock cycle.

The DOB has an even simpler digital form.
The Q-filter of Eq.~(\ref{eq:qfilter}) is realized as a first-order exponential moving average (EMA) filter:
\begin{equation}
 q[n] = (1 - \alpha)\, q[n-1] + \alpha\, x[n]\,,
 \label{eq:ema}
\end{equation}
where $x[n] = K_n\, y[n] - u_\text{out}[n]$ is the raw disturbance estimate and $\alpha$ is the smoothing parameter.
Choosing $\alpha$ as a power of two, $\alpha = 2^{-n}$, eliminates the need for a hardware multiplier: the product $\alpha \cdot x$ becomes a right bit-shift by $n$ positions.
The Q-filter cutoff frequency, in the small-$\alpha$ limit, is then
\begin{equation}
 f_c = \frac{2^{-n}\, f_\text{clk}}{2\pi}\,,
 \label{eq:fc_digital}
\end{equation}
controlled by the single integer parameter $n$.
The entire DOB computation per clock cycle---gain, subtraction, and EMA filtering---reduces to additions and bit-shifts with no floating-point arithmetic, making field-programmable gate array (FPGA) implementation straightforward.
The PID and DOB paths execute in parallel at the full sampling rate, with a fixed pipeline latency that we quantify in Sec.~\ref{sec:setup}.

\begin{figure}[!t]
\centering
\includegraphics[width=\columnwidth]{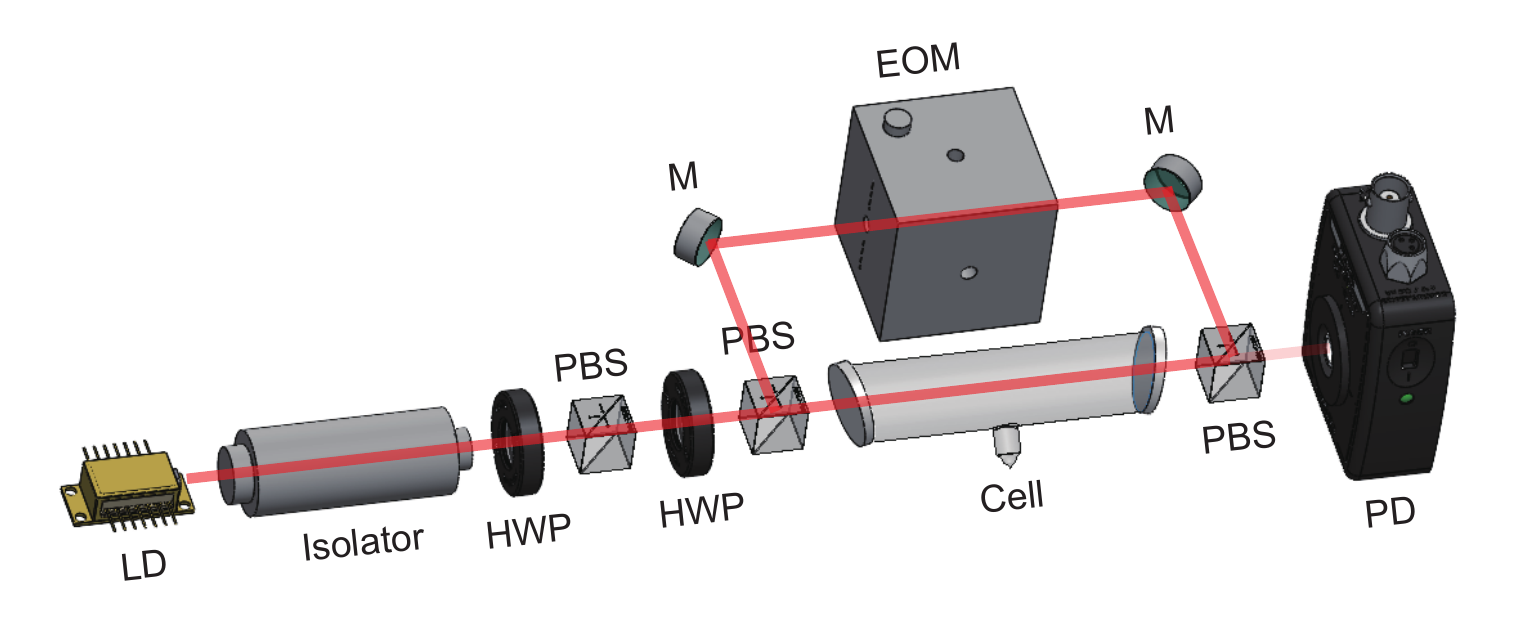}
\caption{Optical layout of the modulation transfer spectroscopy system. The pump beam is phase-modulated at \SI{5}{\mega\hertz} and counter-propagates through the $^{87}$Rb vapor cell with the unmodulated probe beam. LD: laser diode, HWP: half-wave plate, PBS: polarizing beam splitter, M: mirror, EOM: electro-optic modulator, PD: photodetector.}
\label{fig:optical_layout}
\end{figure}

\section{Experimental setup}
\label{sec:setup}

The experimental apparatus consists of three subsystems: the spectroscopy optics for generating the MTS error signal, the FPGA-based digital servo for PID and DOB feedback, and the measurement instruments for characterizing frequency noise and stability.

The spectroscopy system employs modulation transfer spectroscopy~\cite{Shirley1982, McCarron2008, Lee2023} on the $^{87}$Rb D$_2$ transition ($F = 2 \to F' = 3$, $\lambda = \SI{780.24}{\nano\meter}$) in a vapor cell (Thorlabs GC19075-RB87).
The laser source is an external-cavity diode laser (Eagleyard miniECL) driven by a compact low-noise laser controller (Koheron CTL200-1).
A \SI{5}{\mega\hertz} phase modulation is applied to the pump beam via an electro-optic modulator (Qubig PM7-NIR), while the unmodulated probe beam counter-propagates through the cell.
The probe beam power is \SI{150}{\micro\watt} with a $1/e^2$ intensity radius of \SI{3}{\milli\meter}, and the pump beam power is \SI{4.5}{\milli\watt} with a $1/e^2$ intensity radius of \SI{6}{\milli\meter}.
The probe beam is detected by an amplified photodetector (Thorlabs PDA10A2).
The optical layout of the MTS spectroscopy is shown in Fig.~\ref{fig:optical_layout}.

The digital servo system is built on two Red Pitaya STEMlab 125-14 LN platforms, each featuring a Xilinx Zynq-7010 system-on-chip with dual 14-bit \SI{125}{\mega\hertz} analog-to-digital and digital-to-analog converters.
The detector output is routed through a bias tee (Mini-Circuits ZFBT-6GW+) and a low-noise amplifier (Mini-Circuits ZFL-500LN+) to the first unit (RP\#1), which performs modulation and demodulation: a direct-digital-synthesis module generates the \SI{5}{\mega\hertz} carrier that drives the EOM, and a digital lock-in demodulator extracts the dispersive MTS error signal from the detector output.
The demodulated error signal is transmitted from RP\#1 RF~OUT2 to the second unit (RP\#2) via a coaxial cable.

\begin{figure*}[!t]
\centering
\includegraphics[width=0.8\linewidth]{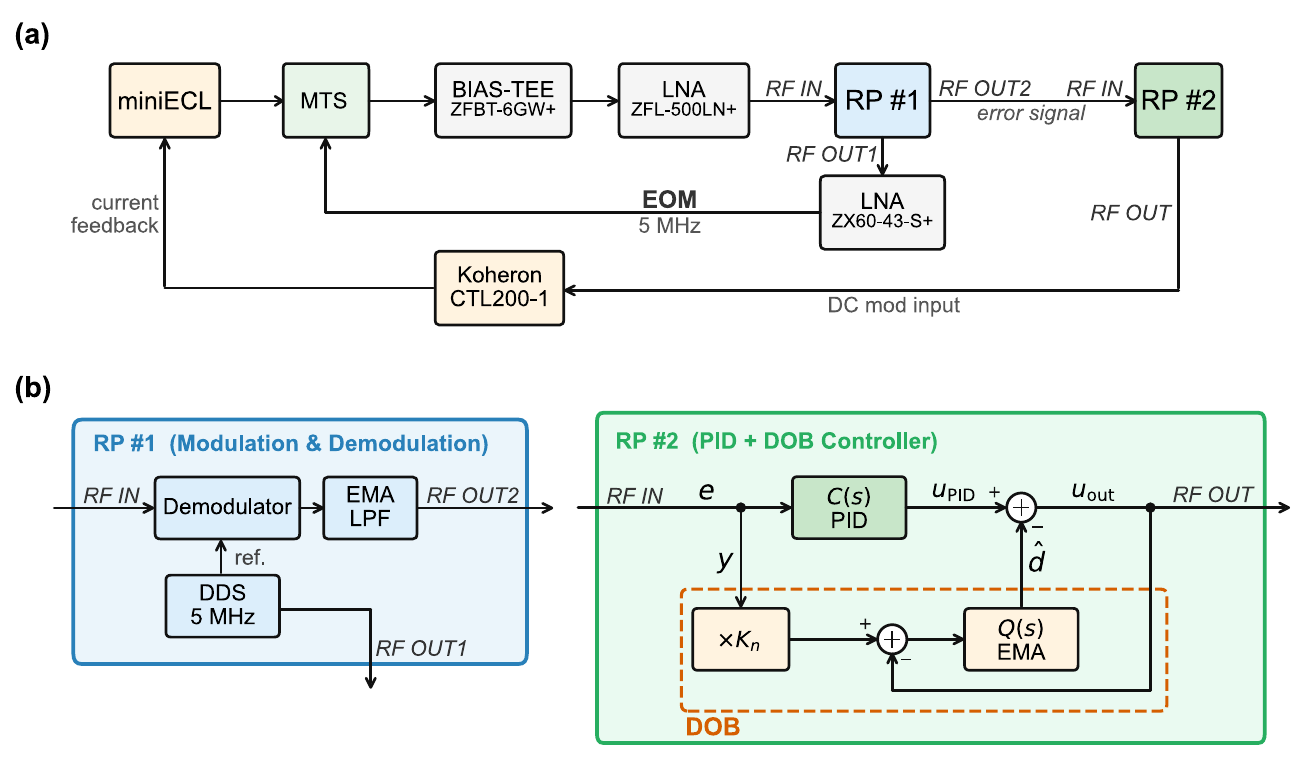}
\caption{Electronic signal flow diagram of the FPGA-based PID+DOB servo system. (a)~External signal path connecting the optical and electronic subsystems. (b)~Internal block diagrams of RP\#1 and RP\#2. The DOB inner loop in RP\#2 implements the simplified form of Eq.~(\ref{eq:d_est_simple}), where the inverse plant model $P_n^{-1}$ of the general DOB architecture (Fig.~\ref{fig:dob_block}) is replaced by the scalar gain $K_n$.}
\label{fig:electronic_layout}
\end{figure*}

Both RP\#1 and RP\#2 implement their respective digital signal-processing functions in custom Verilog RTL. The block diagrams of Fig.~\ref{fig:electronic_layout}(b) are one-to-one graphical representations of these RTLs, with each block corresponding to a Verilog module and the signal arrows corresponding to the synthesized data path. RP\#2 in particular realizes the PID+DOB architecture described in Sec.~II.

The DOB Q-filter is implemented as the EMA of Eq.~(\ref{eq:ema}) with default cutoff $n = 9$, giving $f_c \approx \SI{39}{\kilo\hertz}$. The DOB pipeline (five clock cycles, \SI{40}{\nano\second}) contributes only \SI{0.56}{\degree} of phase lag at $f_c$, well within the robust-stability margin of Shim and Jo~\cite{Shim2009}.
The combined output $u_\text{out} = u_\text{PID} - \hat{d}$ drives the DC current modulation input of the Koheron CTL200-1 laser controller.
The electronic signal flow is illustrated in Fig.~\ref{fig:electronic_layout}.
Comparing the DOB inner loop in Fig.~\ref{fig:electronic_layout}(b) with the general architecture of Fig.~\ref{fig:dob_block}, the inverse plant model $P_n^{-1}(s)$ is replaced by a single gain element $K_n$.
At low frequencies, all analog components (specified above) have cutoffs far exceeding $f_c$ and contribute essentially flat response. The cascaded plant $P(s)$ is therefore dominated by the digital low-pass filters of the lock-in demodulator ($\approx \SI{1.24}{\mega\hertz}$) and the PID EMA pre-filters ($\approx \SI{310}{\kilo\hertz}$), whose cutoffs both lie well above $f_c$. Hence $P(s) \approx P_0$ and $P_n^{-1}(s) \approx K_n$.
The disturbance estimate is then computed as Eq.~(\ref{eq:d_est_simple}), and the DOB sensitivity function $S_\text{DOB}(s) = 1 - Q(s)$ remains approximately valid in this regime.
This design-driven constant-gain approximation, together with the conservative Q-filter bandwidth, replaces the multi-parameter frequency-domain plant identification of conventional DOB designs with a single one-dimensional gain sweep on $K_n$ (demonstrated experimentally in Sec.~\ref{sec:results}).

To characterize the frequency noise PSD, we have recorded the MTS error signal with an FFT signal analyzer (Thorlabs PNA-1).
The analyzer employs 18-bit analog-to-digital conversion across a measurement bandwidth of DC to 3 MHz, with a noise floor below 60 nV/$\sqrt{\text{Hz}}$.
The error signal voltage is converted to frequency units using the discriminator slope, obtained from the linear region of MTS waveforms recorded at different laser currents and calibrated against the manufacturer's specified wavelength tuning coefficient for the Eagleyard miniECL ($d\lambda/dI \approx 0.001$ nm/mA).
For Allan deviation measurements, we have constructed a beat-note setup using two independent, nominally identical MTS-stabilized laser systems.
One beam is frequency-shifted by \SI{100}{\mega\hertz} using a fiber-coupled acousto-optic modulator (Aerodiode 780AOM\_1), and the two beams are combined in a 50:50 fiber coupler (Thorlabs PN780R5A2), with each laser contributing approximately \SI{200}{\micro\watt} of optical power at the coupler input.
The frequency of the beat note is detected by a photodetector (Thorlabs PDA10A2) and measured by a frequency counter (Keysight 53230A) operated in continuous gap-free frequency measurement mode, enabling proper Allan deviation computation with no dead time between gate intervals.
In this mode the gate time is adjustable from \SI{10}{\micro\second} to \SI{1000}{\second} in \SI{10}{\micro\second} steps, with up to $10^6$ samples per acquisition stored in the instrument's internal memory.
We have used two complementary configurations: a short-term measurement with a \SI{1}{\milli\second} gate time and $10^6$ gap-free samples (total duration \SI{1000}{\second}), and a long-term measurement with a \SI{1}{\second} gate time and $2.3 \times 10^4$ gap-free samples (total duration $\sim$\SI{6.4}{\hour}).

\section{Results}
\label{sec:results}

\begin{figure}[!t]
\centering
\includegraphics[width=\columnwidth]{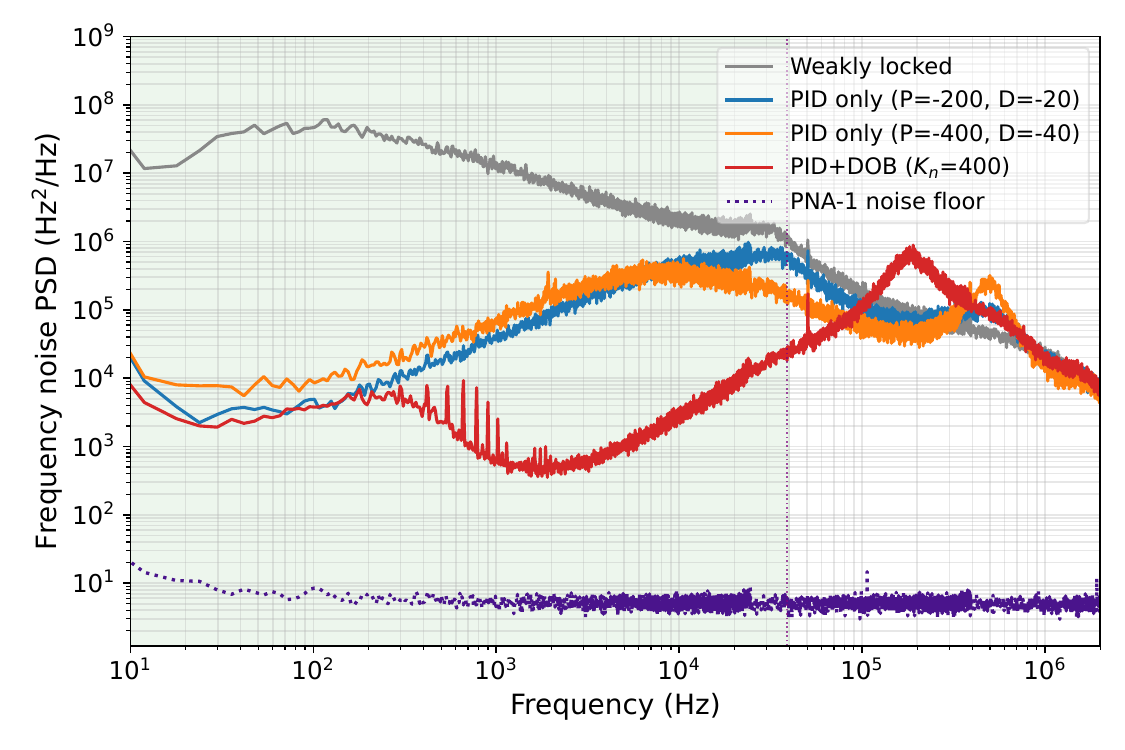}
\caption{Frequency-noise PSD of the MTS error signal for four operating conditions: weakly locked baseline (gray), PID only (blue), high-gain PID only (orange), and PID+DOB (red). See text for the gain settings. The PNA-1 noise floor (purple dotted) shows the signal analyzer's intrinsic limit. The green shaded region marks the integration window (up to \SI{40}{\kilo\hertz}). The dotted vertical line marks the Q-filter cutoff at \SI{39}{\kilo\hertz}. Discrete peaks visible across all traces are residual power-line harmonics.}
\label{fig:psd_compare}
\end{figure}

Figure~\ref{fig:psd_compare} compares the frequency-noise PSD under four operating conditions: a weakly locked baseline (integrator only at minimum gain), approximating the free-running laser; PID only ($K_p = -200$, $K_i = -200$, $K_d = -20$); high-gain PID only ($K_p = -400$, $K_d = -40$); and PID+DOB ($K_p = -200$, $K_i = -200$, $K_d = -20$, $K_n = 400$).
The PID gains were tuned to just below the onset of sustained closed-loop (servo) oscillation. The high-gain trace shows the loop being pushed past this limit into a developed servo bump.

The weakly locked baseline shows a broad frequency-noise pedestal of order $10^6\;\text{Hz}^2/\text{Hz}$ across DC--\SI{100}{\kilo\hertz}.
With PID alone (blue/orange traces), the integrator strongly suppresses noise below \SI{100}{\hertz}, leaving a residual broadband frequency noise that extends up to $\sim$\SI{40}{\kilo\hertz}, peaking near \SI{24}{\kilo\hertz} at the closed-loop servo bump set by the joint action of all PID terms.
Higher PID gains (orange trace) only reshape this residual noise (integrated rms drops only $\sim$\SI{3}{\decibel}) before exciting a developed servo bump that drives the loop into oscillation, so PID alone cannot remove this residual noise without instability. A more elaborate digital-filter design might mitigate this, but at the cost of multi-parameter tuning.

Engaging the DOB---added in parallel without re-tuning the PID---reduces the PSD across this region by nearly two orders of magnitude: the integrated rms frequency noise below \SI{40}{\kilo\hertz} drops by \SI{16.9}{\decibel}, from approximately \SI{140}{\kilo\hertz} (PID only) to approximately \SI{20}{\kilo\hertz} (PID+DOB).
Below $f_c \approx \SI{39}{\kilo\hertz}$ the suppression is maximal. Above $f_c$, traces with and without DOB converge, confirming the low-pass rejection predicted by Eq.~(\ref{eq:sdob}). A relocated servo bump appears near \SI{200}{\kilo\hertz}, above the \SI{40}{\kilo\hertz} integration upper bound. The DOB raises the loop gain below $f_c$, and the combined system's high-frequency response---shaped by the Q-filter rolloff (above $f_c$) and the PID pre-filter cutoff at $\sim$\SI{310}{\kilo\hertz}---develops a new closed-loop resonance between these two scales, observed at $\sim$\SI{200}{\kilo\hertz}.
The signal-analyzer noise floor (PNA-1) is shown for reference.

\begin{figure}[!t]
\centering
\includegraphics[width=\columnwidth]{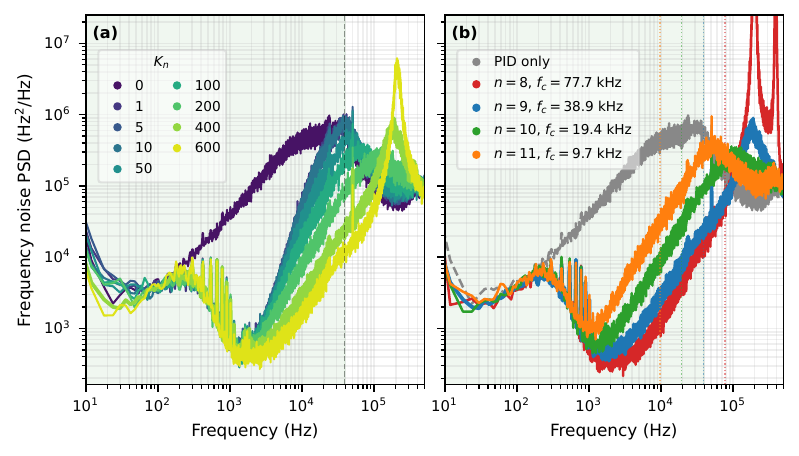}
\caption{Dependence of the frequency-noise PSD on DOB parameters. (a)~DOB gain sweep ($K_n = 0$--$600$, dark to light, viridis colormap) at fixed Q-filter cutoff $n = 9$. The dashed line marks $f_c = \SI{39}{\kilo\hertz}$. (b)~Q-filter cutoff sweep ($n = 8$--$11$) at fixed gain $K_n = 400$. PID only (gray) is shown for reference. Dotted vertical lines mark $f_c$ for each $n$. The green shaded regions indicate the integration window (up to \SI{40}{\kilo\hertz}).}
\label{fig:psd_parameter_sweep}
\end{figure}

The two DOB parameters are characterized by orthogonal sweeps (Fig.~\ref{fig:psd_parameter_sweep}).
At fixed $n = 9$ (Fig.~\ref{fig:psd_parameter_sweep}a), every $K_n$ trace exhibits the characteristic $|S_\text{DOB}| = |1-Q|$ shape predicted by Eq.~(\ref{eq:sdob})---maximal suppression below $f_c$, transparent above---with the suppression depth growing monotonically with $K_n$, from \SI{1.8}{\decibel} ($K_n = 1$) to \SI{19.8}{\decibel} ($K_n = 600$).
At higher $K_n$ the excess above $f_c$ also grows as the strengthened inner-loop response extends toward the PID pre-filter cutoff, signalling the eventual onset of instability.
At fixed $K_n = 400$ (Fig.~\ref{fig:psd_parameter_sweep}b), varying $n = 8$--$11$ shifts the suppression bandwidth from $f_c = \SI{9.7}{\kilo\hertz}$ to \SI{77.7}{\kilo\hertz}, in agreement with Eq.~(\ref{eq:fc_digital}). At the smallest $n$ tested ($n = 8$) the relocated servo bump enters the \SIrange{50}{500}{\kilo\hertz} region, reaching the PID pre-filter cutoff ($\sim$\SI{310}{\kilo\hertz}, Sec.~\ref{sec:setup}), and the loop oscillates.

\begin{figure}[!t]
\centering
\includegraphics[width=\columnwidth]{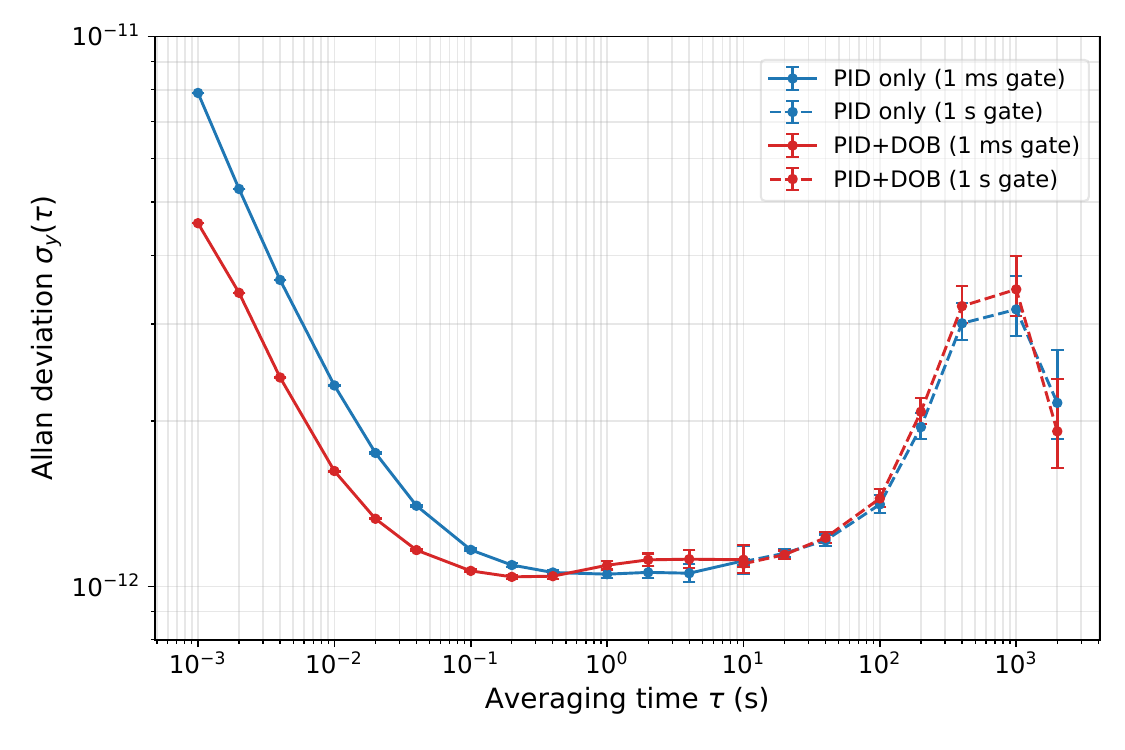}
\caption{Allan deviation of the beat-note frequency. Short-term data (solid lines, \SI{1}{\milli\second} gate) and long-term data (dashed lines, \SI{1}{\second} gate) are shown for PID only (blue) and PID+DOB (red).}
\label{fig:adev}
\end{figure}

Figure~\ref{fig:adev} shows the time-domain stability via the Allan deviation~\cite{Allan1966} $\sigma_y(\tau)$ of the beat note between two independently stabilized laser systems.
With PID only, $\sigma_y(\SI{1}{\milli\second}) = 7.9 \times 10^{-12}$. Engaging the DOB improves this to $4.6 \times 10^{-12}$ (factor of 1.7).
At $\tau = \SI{1}{\second}$ the Allan deviations are $1.05 \times 10^{-12}$ (PID only) and $1.09 \times 10^{-12}$ (PID+DOB).
The DOB's short-term improvement corresponds to the low-frequency PSD reduction shown in Fig.~\ref{fig:psd_compare} (quantitatively cross-validated below), while at long $\tau$ the integrator alone already suppresses low-frequency noise to the asymptotic floor, leaving no residual disturbance for the DOB to address. Both traces share a bump near $\tau \sim \SI{1000}{\second}$ from periodic laboratory temperature fluctuations, common to both configurations.

Finally, we cross-validate by integrating the measured PSD with the standard transfer function~\cite{Allan1987}:
\begin{equation}
 \sigma_y^2(\tau) = \frac{2}{\nu_\text{opt}^2}\int_0^{\infty} S_\nu(f)\, \frac{\sin^4(\pi f \tau)}{(\pi f \tau)^2}\, df.
 \label{eq:adev_from_psd}
\end{equation}
Single-laser predictions are $\sigma_y(\SI{1}{\milli\second}) \approx 1.1 \times 10^{-11}$ (PID only) and $3.7 \times 10^{-12}$ (PID+DOB). For two independent lasers the expected beat-note values are $\sqrt{2}$ times these, i.e., $1.6 \times 10^{-11}$ and $5.2 \times 10^{-12}$.
The measured $\sigma_y(\SI{1}{\milli\second}) = 7.9 \times 10^{-12}$ and $4.6 \times 10^{-12}$ agree with the predictions to within a factor of two, providing internal cross-validation of the DOB-induced reduction.

\section{Conclusion}

We have demonstrated a digital DOB applied to laser frequency stabilization.
A digital DOB, implemented as a first-order EMA Q-filter on a Red Pitaya STEMlab 125-14 LN FPGA, was combined with a conventional PID controller to stabilize a \SI{780.24}{\nano\meter} external-cavity diode laser to the $^{87}$Rb D$_2$ transition via modulation transfer spectroscopy.
The DOB reduced the integrated rms frequency noise below \SI{40}{\kilo\hertz} by \SI{16.9}{\decibel}---from approximately \SI{140}{\kilo\hertz} to \SI{20}{\kilo\hertz}---and improved the short-term Allan deviation at $\tau = \SI{1}{\milli\second}$ from $7.9 \times 10^{-12}$ to $4.6 \times 10^{-12}$, a factor of 1.7.
At $\tau = \SI{1}{\second}$ the Allan deviations are $1.05 \times 10^{-12}$ (PID only) and $1.09 \times 10^{-12}$ (PID+DOB), unchanged by the DOB. At long $\tau$ the integrator alone already suppresses the low-frequency noise to the asymptotic floor, leaving no residual disturbance for the DOB to address. Together with the factor-of-1.7 short-term improvement at $\tau = \SI{1}{\milli\second}$, this confirms the low-pass action of the DOB.

The principal advantage of this approach lies in its simplicity.
Higher-order integral controllers can also achieve similar suppression in a comparable frequency range, but at the cost of multi-parameter tuning and additional pipeline latency. The DOB instead reduces tuning to a single $K_n$ sweep, keeps $|1-Q|$ bounded by unity (no inner-loop amplification), and adds only five clock cycles (\SI{40}{\nano\second}, $\sim$\SI{0.6}{\degree} at $f_c$) of latency, occupying 1836 LUTs and 527 flip-flops with no additional block RAM or DSP slices.
The constant-gain approximation $P_n^{-1} \approx K_n$, enabled by the deliberately flat analog chain, removes the need for explicit plant identification, while the entire DOB computation reduces to additions and bit-shifts with no floating-point arithmetic.

These characteristics are particularly relevant for field-deployable atomic sensors~\cite{Menoret2018, Grotti2018}, where compactness, robustness, and minimal hardware overhead are essential.
Future work will explore higher-order Q-filters for steeper roll-off characteristics and extension to multi-input/multi-output configurations for simultaneous frequency and intensity stabilization.

\begin{acknowledgments}
This work was supported by funding from the Korea government (KASA, Korea Aerospace Administration; grant number RS-2022-00165802) and by the Measurement Technology for Grand National Strategic Industries program funded by the Korea Research Institute of Standards and Science (KRISS 2026-GP2026-0012).
The authors declare no conflicts of interest.
Data underlying the results presented in this paper are available from the corresponding author upon reasonable request.
\end{acknowledgments}

\bibliography{references}

\end{document}